\documentclass[twocolumn,showpacs,superscriptaddress]{revtex4}%
\usepackage{amsmath}
\usepackage{graphicx,epsfig}
\usepackage{dcolumn}
\usepackage{bm}
\usepackage{amsfonts}
\usepackage{amssymb}
\usepackage{graphicx}
\newcommand{\ket}[1]{\left|#1 \right\rangle}
\newcommand{\bra}[1]{\left\langle#1\right|}
\newcommand{\proj}[1]{\ket{#1}\bra{#1}}
\newcommand{\braket}[2]{\left\langle#1|#2\right\rangle}

\newcommand{\half}{\frac{1}{2}}
\newcommand{\tr}{\text{Tr}}
\newcommand{\vis}{\text{v}}

\begin{document}

\title{Experimental Demonstration of Quantum State Multi-meter and One-qubit
Fingerprinting in a Single Quantum Device}

\author{Jiangfeng Du}
\email{djf@ustc.edu.cn}
\address{Hefei National Laboratory for Physical Sciences at Microscale and Department
of Modern Physics, University of Science and Technology of China, Hefei, Anhui
230026, P.R. China}
\address{Department of Physics, National University of Singapore, 117542 Singapore}
\author{Ping Zou}
\address{Hefei National Laboratory for Physical Sciences at Microscale and Department
of Modern Physics, University of Science and Technology of China, Hefei, Anhui
230026, P.R. China}
\author{Daniel K. L. Oi}
\address{Centre for Quantum Computation, DAMTP, University of Cambridge, Wilberforce
Road, Cambridge CB3 0WA , U.K.}
\author{Xinhua Peng}
\address{Universit\"{a}t Dortmund, Fachbereich Physik, 44221 Dortmund, Germany}
\author{L.C.Kwek}
\address{Department of Natural Sciences, National Institute of Education, Nanyang
Technological University, 1 Nanyang Walk, Singapore 637616.}
\author{C.H. Oh}
\address{Department of Physics, National University of Singapore, 117542 Singapore}
\author{Artur Ekert}
\address{Centre for Quantum Computation, DAMTP, University of Cambridge, Wilberforce
Road, Cambridge CB3 0WA , U.K.}

\begin{abstract}
  We experimentally demonstrate in NMR a quantum interferometric multi-meter
  for extracting certain properties of unknown quantum states without resource
  to quantum tomography. It can perform direct state determinations,
  eigenvalue/eigenvector estimations, purity tests of a quantum system, as well
  as the overlap of any two unknown quantum states. Using the same
  device, we also demonstrate one-qubit quantum fingerprinting.
\end{abstract}

 \maketitle

Quantum information processors exploit the quantum features of superposition
and entanglement for applications not possible in classical devices, offering
the potential for significant improvements in the communication and processing
of information. Many quantum circuits have been developed for quantum
information processing, both for efficient algorithms and for secure
communication. Of these, a versatile circuit is the so-called `scattering'
circuit~\cite{kitaev,cleve,llyod,maquel,buhrman,ekert,horodecki}.  Versions of
this circuit play a crucial role in many quantum algorithms exhibiting marked
improvements over the best classical counterparts. For example, it occurs in
Kitaev's solution to the Abelian stabilizer problem~\cite{kitaev}, in the
analysis of quantum algorithms~\cite{cleve}, for finding approximate
eigenvalues of certain Hamiltonians~\cite{llyod}, demonstrating tomography and
spectroscopy as dual forms of quantum computation~\cite{buhrman}, quantum
fingerprinting~\cite{buhrman}, direct estimations of linear and nonlinear
functions of a quantum state~\cite{ekert}, as well as direct detection of
quantum entanglement\cite{horodecki}.

Correspondingly, much effort has been made to physically realize quantum
devices in many different physical systems. Nuclear magnetic resonance (NMR)
has been the first to demonstrate non-trivial quantum algorithms with small
numbers of qubits~\cite{chuang}. In this letter, we demonstrate the experimental
implementation of the above quantum device using three qubits and nuclear magnetic
resonance (NMR) techniques. We demonstrate its application
to direct state determination and eigenvalue/eigenvector estimation of
unknown qubit states, estimation of the overlap of any two states, and
the implementation of a qubit fingerprinting scheme~\cite{beaudrap}.

\begin{figure}[ptb]
\includegraphics[width=0.45\textwidth]{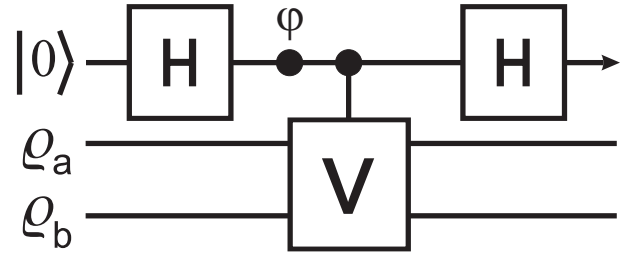}
\caption{The `Scattering Circuit'. The top line represents an ancilla qubit,
  initialised in the state $\ket{0}$, acting as a ``probe particle''. The lower
  two lines represent two physical systems of interest. A controlled-$V$
  operation is applied between two Hadamard gates and a phase shift gate. A
  measurement of the probe in the $\{0,1\}$ basis reveals the overlap of the
  two input states.}
\label{fig:circuit}
\end{figure}

The network of the quantum device is shown in Fig.\ref{fig:circuit}. The
Hadamard gate ($H$) maps
$\ket{b}\mapsto\frac{1}{\sqrt{2}}(\ket{0}+(-1)^{b}\ket{1}),b\in\{0,1\}$, and
the phase shift gate $\varphi=e^{-i\varphi\sigma_{z}/2}$ rotates
the qubit by the angle $\varphi$ about the $z$ axis. The controlled-$V$ gate is
the controlled-SWAP gate (also called quantum Fredkin gate), which acts
trivially if the control qubit is in state $\ket{0}$ and swaps the states of
the lower systems if the control qubit is in the state $\ket{1}$, where
$V\ket{\alpha}_{a}\ket{\beta}_{b}=\ket{\beta}_{a}\ket{\alpha}_{b}$ for all pure
states $\ket{\alpha}$\ and $\ket{\beta}$. With general two input states
$\varrho_{a}$ and $\varrho_{b}$, the reduced density matrix of the ancilla
qubit at the end will have the form
\begin{equation}
\begin{pmatrix}
\frac{1}{2}(1+\cos\varphi\tr\left[\varrho_{a}\varrho_{b}\right]) &
\frac{1}{2}i\sin\varphi\tr\left[\varrho_{a}\varrho_{b}\right]\\
-\frac{1}{2}i\sin\varphi\tr\left[\varrho_{a}\varrho_{b}\right] &
\frac{1}{2}(1-\cos\varphi\tr\left[\varrho_{a}\varrho_{b}\right])
\end{pmatrix}.
\label{finalo1}
\end{equation}
The probabilities of measuring the ancilla qubit are
$P_{0}=\frac{1}{2}(1+\cos\varphi\tr\left[\varrho_{a}\varrho_{b}\right])$ in
state $\ket{0}$ and
$P_{1}=\frac{1}{2}(1-\cos\varphi\tr\left[\varrho_{a}\varrho_{b}\right])$ in
state $\ket{1}$. The difference of these two probabilities is $\left\vert
  P_{0}-P_{1}\right\vert=\left\vert\cos\varphi\tr\left[\rho_{a}\rho_{b}\right]\right\vert$,
which can easily be measured on the NMR interferometer. With $\varphi=0$, it is
the visibility $\vis=\tr\left[\varrho_{a}\varrho_{b}\right]$.

We implemented the circuit in liquid-state NMR using as qubits the three
spin-$\half$ carbon nuclei in a $^{13}C$-labeled sample of alanine
$NH_{3}^{+}-C^{^{\alpha}}H(C^{^{\beta}}H_{3})-C^{\prime}OOH$
in deuterated water. $C^{\prime}$ forms the ancilla system which is
denoted as control qubit-1, $C^{\alpha}$ and $C^{\beta}$ form the target
systems that are denoted as qubit-2 and qubit-3 (top, middle and bottom lines
in Fig.\ref{fig:circuit}). With decoupling of the protons, this spin system
exhibits a weakly coupled spectrum corresponding to the Hamiltonian
$H=\sum_{i=1}^{3} \omega_i\frac{\sigma_z^i}{2}+
\frac{\pi}{2}\left(J_{i,i+1}\sigma_x^i\sigma_z^{i+1}\right)$ where $\sigma_{k}$
are rescaled Pauli matrices, $\frac{\omega_{i}}{2\pi}$ are Larmor frequencies
and $J_{ij}$ are spin-spin coupling constants.  The experiments were carried
out at the National Institute of Sciences of Nanyang Technological University
on a Bruker Avance $DMX400$ spectrometer in a field of roughly $9.4T$ equipped
with a $5mm$ probe. The frequency shifts of the other carbons with respect to
the third are $12609.6Hz$ for the first and $-3455.7Hz$ for the second, while
the coupling constants are $J_{12}=-1.2Hz$, $J_{23}=35Hz$, and $J_{13}=54Hz$.
Longitudinal relaxation times ($T_1$) for all three spins exceeded $1.5s$,
while the transverse relaxation times ($T_{2}$) were at least $420ms$.

Experimentally, the whole process of the device is demonstrated in three
steps:

(1) Prepare the input state $\proj{0}\otimes\varrho_{a}\otimes\varrho_{b}$: As
is standard, the density matrix of a spin-$\half$ particle can be written in
terms of the Bloch vector $\vec{r}$ and the Pauli matrices
$\vec{\sigma}=\{\sigma_{x},\sigma_{y},\sigma_{z}\}$ as $\rho
=\half\left(I+\vec{r}\cdot\vec{\sigma}\right)$ (where $I$ is the unit matrix).
The length of the Bloch vector $r$ gives a measure of purity (pure $r=1$ to
maximally mixed $r=1$). In our experiments, we first use the spatial labelling
method~\cite{cory} to prepare the effective pure state
$\rho_{pp}=\proj{000}=\frac
{\left(I+\sigma_{z}^{1}\right)}{2}\otimes\frac{\left(I+\sigma_{z}^{2}\right)}{2}
\otimes\frac{(I+\sigma_{z}^{3})}{2}$. To prepare arbitrary mixed states
$\rho_{a}$ and $\rho_{b}$ from $\rho_{pp}$, we adopt the similar method in
previous experiments where both unitary operators and non-unitary operators are
used~\cite{duprl}.

For example, to create the mixed states $\varrho_{a}\otimes\varrho_{b}=\left(
  \frac{I}{2}+\frac{\sqrt{2}}{2}\sigma_{z}^{2}\right)\otimes\left(\frac
  {I}{2}+\frac{1}{2}\sigma_{x}^{3}\right)$ from $\rho_{pp}$, two spin-selective
radio frequency (RF) pulses, $e^{-i\frac{\pi}{8}\sigma_{x}^{2}}$ and
$e^{-i\frac{\pi}{6}\sigma_{x}^{3}}$, are applied on qubit-2 and qubit-3
respectively. They transform
$\frac{\left(I+\sigma_{z}^{2}\right)}{2}\otimes\frac{(I+\sigma_{z}^{3})}{2}$ to
the state
$\left(\frac{I}{2}-\frac{\sqrt{2}}{2}\sigma_{y}^{2}+\frac{\sqrt{2}}{2}\sigma_{z}^{2}\right)
\otimes\left(\frac{I}{2}-\frac{\sqrt{3}}{2}\sigma_{y}^{3}+\frac{1}{2}\sigma_{z}^{3}\right)$.
A pulsed field gradient (PFG) in $z$-direction is then applied to dephase
off-diagonal elements of the density matrix leading the state
$\left(\frac{I}{2}+\frac{\sqrt{2}}{2}\sigma_{z}^{2}\right)\otimes
\left(\frac{I}{2}+\frac{1}{2}\sigma_{z}^{3}\right)$.  Finally, by applying
another spin selective RF pulse $e^{-i\frac{\pi}{4}\sigma_{y}^{3}}$, the
desired state is prepared. Therefore, the Bloch lengths of spin $2$ and $3$
have been shortened to purities $r_{a}=\frac{\sqrt{2}}{2}$ and
$r_{b}=\frac{\sqrt{3}}{2}$ by non-unitary operations, while the orientation of
the spin vector has been rotated to the desired direction by unitary
operations. Hence, we can generate the state of each spin with any length and
orientation in the Bloch sphere. In the experiments, all the spin selective
pulses are Gaussian-shaped and have the same pulse duration $0.866ms$,
with the rotation magnitude determined by the pulse power.

(2) Application of the circuit: The single qubit gates $H$ and $\varphi$ gate
are easily realized in NMR~\cite{chuang}. However, it is difficult to directly
realize the three-qubit Fredkin gate~\cite{fredkin}. In principle, the gate can
be decomposed into one and two-qubit gates~\cite{barenco}. Chau and Wilczek
gave a construction using six specific gates~\cite{chau}, Smolin and DiVincenzo
then showing that five two-qubit gates were optimal~\cite{smolin}. However, it
is still difficult to realize the three-qubit Fredkin gate precisely in
practice. In NMR, one efficient choice is to directly construct multi-qubit
quantum gates by using low-power RF pulses on a single multiplet-component
(transition-selective excitation or TSE). This takes advantage of the full
complexity of the internal Hamiltonian and uses the scalar coupling between
pairs of spins to construct multi-qubit logic gates, which operate on many
qubits simultaneously~\cite{dupra}.

We implement the circuit by the pulse sequence (left to the right)
\begin{equation}
R_{\varphi}^{1}\left(\frac{\pi}{2}\right)
-TP1-TP2-TP3
-R_{\delta}^{1}\left(\frac{\pi}{2}\right),
\label{pulses}
\end{equation}
where $R_{\varphi}^{1}(\frac{\pi}{2})=e^{-i\frac{\pi}{4}(\sigma_{x}\cos
  \varphi+\sigma_{y}\sin\varphi)}$ denotes a $\pi/2$ selective pulse that acts
on qubit-$1$ about the axis $\widehat{x}\cos\varphi+\widehat{y}\sin\varphi$,
combining the first Hadamard and the $\varphi$ gate in Fig.~\ref{fig:circuit}.
The next three transition pulses perform a modified Fredkin gate which induces
a nontrivial phase factor~\cite{xue}. The duration of each Gaussian-shaped TSE
pulses were $73.5ms$ for sufficient selectivity in the frequency domain without
disturbing the nearest line (see~\cite{peng} for detailed analysis of TSE
pulses).

\begin{figure}[ptb]
\includegraphics[width=0.45\textwidth]{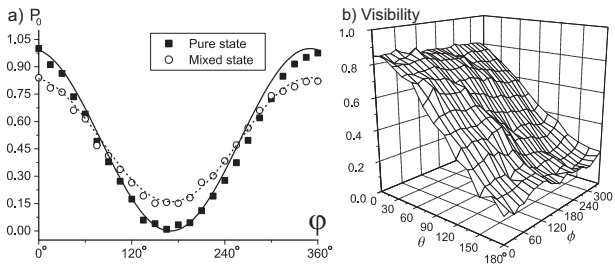}
\caption{a) The interference pattern for both pure and mixed state as function
  of the angle $\varphi$ of phase gate.  The boxes and circles are experimental
  results when both input states $\{\varrho_a,\varrho_b\}$ are equal to
  $\frac{I+\sigma_{z}}{2}$ or $\frac{I}{2}+\frac{\sqrt{3}\sigma_{z}}{4}$
  respectively. The solid and dotted curves correspond to the theoretical
  interference pattern
  $P_{0}=\frac{1}{2}(1+\cos\varphi\cdot\tr\left[\varrho_{a}\varrho_{b}\right])$.
  b) The experimental visibilities $\tr[\varrho_{a}\varrho_{b}]$ measured when
  the pure state $\varrho_{a}(\theta,\phi)$ is scanned over the Bloch sphere by
  changing the angle $\theta$ and $\phi$. The second input state, $\varrho_b$
  is fixed as $\varrho_{b}=\frac{I}{2}+\frac{\sqrt{2}\sigma_{z}}{4}$.}
\label{fig:mixed}
\end{figure}

(3) Measure the ``probe'' qubit-1. The final reduced density
matrix of qubit-1 is
\begin{eqnarray}
\varrho_{1}
=\left(
\begin{array}{cc}
\varrho_{00} & \varrho_{01}\\
\varrho_{10} & \varrho_{11}
\end{array}
\right)
&=&(\varrho_{01}+\varrho_{10})\frac{\sigma_{x}^{1}}{2}
  +(\varrho_{01}-\varrho_{10})i\frac{\sigma_{y}^{1}}{2}\nonumber\\
&+&(\varrho_{00}-\varrho_{11})\frac{\sigma_{z}^{1}}{2}
   +(\varrho_{00}+\varrho_{11})\frac{I}{2}.
\end{eqnarray}
We will obtain the difference between the probabilities of finding qubit-1 in
state $\ket{0}$ and $\ket{1}$ when the coefficient of
$\frac{\sigma_{z}^{1}}{2}$ is measured. We first apply a PFG to remove the
non-diagonal part of the density matrix, then a pulse
$R_{\frac{\pi}{2}}^{1}(\frac{\pi}{2})$. The state of qubit-1 is now
$(\varrho_{00}-\varrho_{11})\frac{\sigma_{x}^{1}}{2}+\frac{1}{2}(\varrho_{00}+\varrho_{11})I$.
Since the identity matrix in NMR is not observable, the integral area of peaks
of qubit-1 is now proportional to
$(\varrho_{00}-\varrho_{11})$=$\cos\varphi\tr\left[\varrho_{a}\varrho_{b}\right]$.
Practically, this is measured by integrating the entire multiplet and adding
together the signals arising from the four components of qubit-1.
Fig.\ref{fig:mixed}a compares interference patterns, from theory~\cite{ekert}
and experiment, for both the pure and mixed states. Note that the additional
phase factor induced by the three TSE pulses has no effect in testing the
interference pattern since both the Bloch vectors of the inputs $\varrho_{a,b}$
are in the $z$-direction, the angle $\delta$ of right pulse shown in
Eq.(\ref{pulses}) is set to $\frac{\pi}{2}$. It can be seen that the phase
shift due to the SWAP operator is zero, and the visibility decreases when the
input state changes from being pure $\left(r=1\right)$ to mixed
$(r=\sqrt{3}/2)$.

We now implement a quantum multimeter and perform quantum fingerprinting using
the circuit. A multimeter can be used to estimate the properties of a quantum
state, or compare two quantum states, which are basic quantum processing tasks.
Quantum fingerprinting is a proof-in-principle of an exponential
quantum/classical gap for the equality problem in the simultaneous message
passing model of communication complexity~\cite{buhrman,beaudrap,massar}. In
both applications, to cancel the effect of the additional phase factor induced
by three TSE pulses, we set the $\varphi=\pi/4$ and $\delta=3\pi/2$ in
Eq.(\ref{pulses}).

\textit{Multimeter for an unknown state:} We detect an unknown qubit state of
$\varrho_b$ and find its eigenvalue and eigenstate by using some particular
known states as input state $\varrho_{a}$. We show below how this can be done
by preparing suitable initial states $\varrho_{a}$.

We set input $\varrho_{b}$ as the unknown state, and separately prepare three
comparison states $\proj{\psi_a}\in\left\{\frac{I+\widehat{\sigma}_{x}}{2},
\frac{I+\widehat{\sigma}_{y}}{2},\frac{I+\widehat{\sigma}_{z}}{2}\right\}$ as
$\varrho_{a}$. Therefore the visibility of qubit-1 in each run
contains information of the unknown state $\varrho_{b}$, which corresponds to
the expectation value
$\left\langle\sigma_{bi}\right\rangle=\frac{1+r_{bi}}{2},i=\{x,y,z\}$. Hence we
can determine the density matrix $\varrho_{\text{exp}}$ of the unknown state
from these three values. Experimentally, we test many ``unknown quantum'' states,
evaluating the performance using the Uhlmann fidelity~\cite{U1976},
$\mathcal{F}\left[\varrho_{b},\varrho_{\exp}\right]=
\tr\left[\sqrt{\sqrt{\varrho_{b}}\varrho_{\exp}\sqrt{\varrho_{b}}})\right]^{2}$,
which gave an average fidelity
$\mathcal{F}\left[\varrho_{b},\varrho_{\text{exp}}\right]=0.98\pm0.01$.

By replacing above three states $\ket{\psi_a}$, we can estimate eigenvalues and
eigenvectors of $\varrho_{b}$. We scan
$\varrho_{a}=\frac{I+\vec{r}_{a}\vec{\sigma}}{2}$,
$\vec{r}_{a}=(\cos\theta\cos\phi,\cos\theta\sin\phi,\sin\theta)$, with $\theta$
ranging from $0^{\circ}$ to $180^{\circ}$ in $15^{\circ}$ steps, and $\phi$
from $0^{\circ}$ to $345^{\circ}$ with $15^{\circ}$ steps. We then measure
$\tr\left[\varrho_{a}\varrho_{b}\right]$, and find the minimum and maximum..
The two extrema are the eigenvalues of the unknown state $\varrho_{b}$, and the
corresponding input states are its eigenvectors. Fig.\ref{fig:mixed}b shows
the estimation of eigenvalues and eigenvectors of
$\varrho_{b}=\frac{I}{2}+\frac{\sqrt{2}\sigma_{z}}{4}$. The experimentally
determined eigenvalues are $\{0.833,0.182\}$ for eigenvectors $\ket{0}$
and $\ket{1}$ respectively, which compare well with the ideal results,
$\{\half\pm\frac{\sqrt{2}}{4}\}=\{0.854,0.146\}$.

\textit{Multimeter for two unknown states:} We can also compare two general
quantum states. For pure states $\varrho_{a}=\proj{\alpha}$ and
$\varrho_{b}=\proj{\beta}$, the visibility of qubit-1 gives
$\tr\left[\varrho_{a}\varrho_{b}\right]=\left|\braket{\alpha}{\beta}\right|^{2}$,
i.e. the orthogonality of $\ket{\alpha}$ and $\ket{\beta}$.
For mixed states the visibility provides the measure of overlap
$\tr\left[\varrho_{a}\varrho_{b}\right]$ of $\varrho_{a}$ and $\varrho_{b}$. If
$\varrho_{a}=\varrho_{b}=\varrho$ then
$\tr\left[\varrho_{a}\varrho_{b}\right]=\tr\left[\varrho\right]^{2}$, which is
the purity of $\varrho$.

\begin{figure}[ptb]
\includegraphics[width=0.45\textwidth]{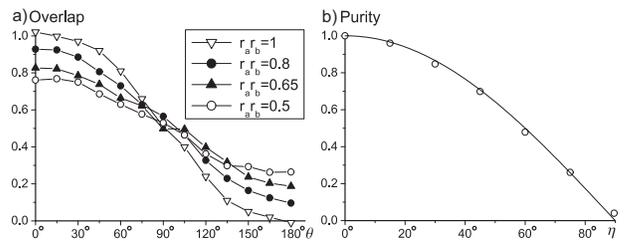}
\caption{a) Overlap as a fuction of $\theta$, the angle between state vectors,
  with different purities $r_{a}$ and $r_{b}$. Four sets of experiments are
  shown in the figure, distinguished by different purities of the two states
  $\left( r_{a},r_{b}\right) =\{\left( 1,1\right) ,\left( 1,0.8\right) ,\left(
    0.81,0.81\right) ,\left( 1,0.5\right) \}$. b)
  Experimental determined purity as a function of $\eta$. The circles
  correspond to experimental results, and the curve corresponds to theory:
  $r=\cos\eta$.}
\label{fig:overlappurity}
\end{figure}

We prepare various states $\varrho_{a}$ and $\varrho_{b}$ as the inputs,
applying the circuit and measuring their corresponding visibilities, i.e., the
overlap
$\tr\left[\varrho_{a}\varrho_{b}\right]=\frac{1+r_{a}r_{b}\cos\theta}{2}$,
where $r_{a,b}$ are the lengths of the Bloch vectors $\vec{r}_{a,b}$, and
$\theta$ is the mutual angle between them. Fig.\ref{fig:overlappurity}a shows the
experimental performance to compare the two unknown quantum states
$\varrho_{a}$ and $\varrho_{b}$. The device successfully gives the direct
estimation of the overlap of two unknown states.  Moreover, for two pure
states, i.e., $r_{a}=r_{b}=1$, the experimental results present the direct
measure of orthogonality.

We can also directly estimate the purity of a mixed state.  We begin by
preparing $\varrho_{a}$ and $\varrho_{b}$ in the same mixed state
$\half(I+r\sigma_{z})$, where $r=\cos\eta$ ($\eta=\frac{n\pi}{12},n=0,1,\ldots,6$)
is the purity. For each mixed state, the visibility
$\tr\left[\rho^{2}\right]=\frac{1+r^{2}}{2}$ is measured from which $r$ can
easily be extracted (Fig.\ref{fig:overlappurity}b).

\textit{Quantum Fingerprinting}: Finally, we demonstrate quantum
fingerprinting~\cite{buhrman}, for which Beaudrap~\cite{beaudrap} recently
defined and presented one-qubit versions which outperform any classical one-bit
schemes. We implement the scheme from~\cite{beaudrap} as follows:

Alice and Bob use the same set of pure states
$\left\{\ket{\phi_\sigma}\right\}_{\sigma\in S}$ as fingerprints. In
particular, we set $\left\vert S\right\vert=6$ and
$\left\{\ket{\phi_\sigma}\right\}_{\sigma\in S}=\{\ket{\pm x},\ket{\pm
  y},\ket{\pm z}\}$.
The absolute value of the inner product of any two distinct states does not
exceed $|\braket{\alpha}{\beta}|\le\delta=\frac{1}{\sqrt{2}}$.

Alice and Bob send to a referee their fingerprints $\ket{\phi_\alpha}$ and
$\ket{\phi_\beta}$ randomly selected from $S$. The referee then needs to
distinguish between the cases $\alpha=\beta$ and $\alpha\ne\beta$. The referee
puts the two states as the inputs $\{\varrho_a,\varrho_b\}$ of the circuit,
from which measurement of the first qubit gives the visibility
$\tr\left[\varrho_{a}\varrho_{b}\right]=
\left\vert\braket{\phi_\alpha}{\phi_\beta}\right\vert ^{2}$.  From measurements
of all 36 combinations, we obtain a maximum overlap of fingerprints of $0.54$
when $\alpha\neq\beta$. Hence, we obtain the experimental one-side error
$0.77$, while the theoretical one-side error is $\frac{1+\delta^{2}}{2}=0.75$.
The error probability is $1$ for any classical one-bit fingerprinting with
one-sided error.

The observable errors in the experiment come from pulse imperfections, both in
the TSE and spin selective pulses, variability over time of the measurement
process, and RF field inhomogeneity. Since the $T_{2}$ relaxation times of the
spins varied from $0.42-0.8s$ compared to the experiment duration of $0.23s$,
the net loss of magnetization due to relaxation is not negligible. This effect
have been reduced by renormalizing the integration of the spectra during the
measurement stage.

In conclusion, we have used a three-spin system and liquid-state NMR to
demonstrate a proof-in-principle quantum device, and we also test several
potential applications in quantum information, both quantum computation and
quantum communication complexity, in this single quantum device. The
implemented quantum circuit forms the basis for many interesting and quantum
information tasks. The experimental results show good agreement with the
theoretical predictions.

\textit{Note added.} After completing this work, it has come to our knowledge
that Rolf T. Horn \textit{et al.}~\cite{sanders} have demonstrated the
single-qubit quantum fingerprinting using linear optics.

This project was supported by Temasek Project in Quantum
Information Technology (Grant No. R-144-000-071-305). We also
thank supports from the National Nature Science Foundation of
China (Grant No. 10075041), Funded by the National Fundamental
Research Program (2001CB309300) and National Science Fund for
Distinguished Young Scholars (Grant No. 10425524). DKLO
acknowledges support from EU grants RESQ (IST-2001-37559) and
TOPQIP (IST-2001-39215), Fujitsu, and Sidney Sussex College
Cambridge.

\end{document}